\documentclass[aps,prb,11pt,footinbib, showpacs,showkeys]{revtex4}
\usepackage{graphicx}
\usepackage{epsfig}
\usepackage{dcolumn}
\usepackage{bm}
\usepackage{amsmath}
\usepackage{amsbsy}
\usepackage{amssymb}
\usepackage{color}
\usepackage{verbatim}

\newcommand{\bra}[1]{\ensuremath{\langle{#1}|\,}}
\newcommand{\ket}[1]{\ensuremath{\,|{#1}\rangle}}

\begin{document}

\title{Photoexcitation of graphene with twisted light}
\author{M.\ B.\ Far\'ias}
\author{G.\ F.\ Quinteiro}
\author{P.\ I.\ Tamborenea}
\affiliation{Departamento de F\'{\i}sica e IFIBA, FCEN,
Universidad de Buenos Aires, Ciudad Universitaria, Pab.\ I,
C1428EHA Buenos Aires, Argentina}


\begin{abstract}
We study theoretically the interaction of twisted light with graphene.
The light-matter interaction matrix elements between the tight-binding 
states of electrons in graphene are determined near the Dirac points.
We examine the dynamics of the photoexcitation process by posing the
equations of motion of the density matrix and working up to second order
in the field.
The time evolution of the angular momentum of the photoexcited electrons
and their associated photocurrents are examined in order to elucidate
the mechanisms of angular momentum transfer.
We find that the transfer of spin and orbital angular momentum from light 
to the electrons is more akin here to the case of intraband than of 
interband transitions in semiconductors, due to the fact that the two 
relevant energy bands of graphene originate from the same atomic orbitals. 
\end{abstract}

\pacs{78.67.Wj, 42.50.Tx}
\keywords{graphene, twisted light, orbital angular momentum}

\maketitle


\section{Introduction}

Due to its low dimensionality and particular crystalline structure, graphene 
presents an unusual semi-metallic behavior, and its low-energy excitations 
behave as massless Dirac fermions.\cite{cas-gui-per-nov-gei,abe-apa-ber} 
Because of this, graphene shows unusual transport properties, like
an anomalous quantum hall effect\cite{IQHE} and Klein 
tunnelling.\cite{sta-hua-gol,you-kim}
Its optical properties are also peculiar: despite being one-atom thick, 
graphene absorbs a significant amount of white light, and its transparency 
is governed by the fine structure constant, usually associated with quantum
electrodynamics rather than condensed matter physics.\cite{opac,opacexp}

In parallel to these discoveries, a new branch of optics, the study of 
phase-structured light, developed vigorously in the last twenty years. 
The generation and applications of light carrying orbital angular momentum 
(OAM) or \textit{twisted light} (TL) gained great attention after the seminal 
work of Allen \textit{et al.},\cite{all1992} who showed that light carrying 
an integer amount of \textit{orbital} angular momentum  ($\hbar \ell$, 
with $\ell$ an integer) may be generated in the laboratory using 
conventional laser beams. 
Current efforts in this field are directed, on the one hand, to the 
understanding and generation of twisted light beams, and, on the other hand, 
to the study of interaction with mesoscopic particles, atoms and molecules, 
and Bose-Einstein condensates.\cite{libro} 
Only recently the interaction with solid-state systems has been studied, 
subject that appears to be a promising field of research and technology. 
Theoretical works study the interaction of TL with various semicondunctor 
systems.\cite{qui2009a,qui2009b,qui2010,and-ryu-cla,teoria2} 
Also experiments on the field are starting to be conducted.\cite{exp1,exp2}

The interaction of graphene with light has been studied theoretically 
with different approaches, for instance by the calculation of the opacity 
and the optical conductivity,\cite{opacexp} or control of photocurrents.
\cite{tomanek}
The study of the interaction of graphene with light carrying OAM is 
interesting by itself, since it leads to questions rarely formulated in works 
studying the optical properties of graphene. 
Since the twisted light has orbital angular momentum, one may expect a 
transfer of OAM from the photons to the electrons in graphene. 
However, the analysis is complicated by the fact that the low-lying 
excitations of graphene are Dirac fermions, whose OAM is not well-defined.
Nevertheless, there is another angular momentum, known as 
\textit{pseudospin} associated with the honeycomb lattice of graphene, 
and the total angular momentum (orbital plus pseudospin) is conserved. 
This variable has been broadly discussed in graphene-related literature, 
\cite{pseudospin1} but it has usually been left aside in works studying 
interaction with light.

In this work, we study theoretically the interaction of graphene with 
twisted light and calculate relevant physical magnitudes: the 
photo-induced electric currents and the transfer of angular momentum.
To this end, we first obtain the light-matter matrix elements for twisted 
light and graphene, and then use quantum-kinetic equations of motion to 
obtain the time evolution of physical observables.\cite{sch-weg,
win-kno-mal,fer-via-blu-per-per-cas,fer-per-rib-sta}

The paper is organized as follows.
In Section \ref{sec:states} we describe the electronic states of graphene.
Next, in Section \ref{sec:Hint} we study the light-matter interaction
Hamiltonian.
Section \ref{sec:em} contains the calculation of the interaction
matrix elements.
In Section \ref{sec:hem} we write the equations of motion for the density
matrix and study them in the low excitation regime.
Relevant physical quantities are examined in Section \ref{sec:epm}.
Finally, conclusions are presented in the last Section.

\section{Low-excitation states in graphene}
\label{sec:states}

\begin{figure}[h]
\centering
\includegraphics[scale=0.8]{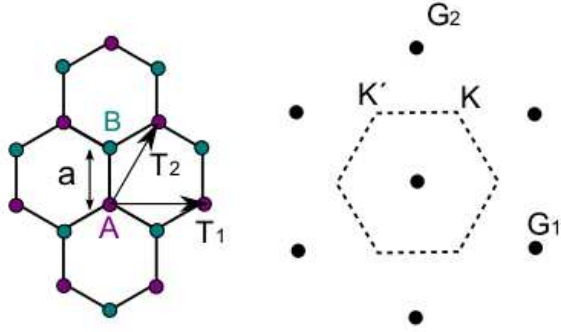}
\caption{Crystal structure of graphene in the direct (left) and reciprocal 
  (right) space. 
  \textbf{T}$_1$ and \textbf{T}$_2$ are the primitive vectors of the 
  Bravais lattice, and \textbf{K} and \textbf{K'} are the corners of the 
  first Brillouin zone (Dirac Points).}
\label{fig:grafeno} 
\end{figure}

In Fig.\ \ref{fig:grafeno} (left) we show the crystalline structure of
graphene: a honeycomb lattice formed by carbon atoms separated by a distance 
$a \approx 1.42 \, \text{\AA}$.
The first Brillouin zone is shown in Fig.\ \ref{fig:grafeno} (right). 
In the tight-binding model (with nearest-neighbour hopping $t$) the 
Hamiltonian matrix is given by

\begin{equation}
H_0(\textbf{k})=t\left(
\begin{array}{c c}
0 & \begin{array}{c}
1+e^{-i\textbf{k}\cdot \textbf{T}_2}+e^{-i \textbf{k}\cdot 
(\textbf{T}_2-\textbf{T}_1)}\end{array}\\
\begin{array}{c} 1+e^{i\textbf{k}\cdot \textbf{T}_2}+e^{i \textbf{k}\cdot 
(\textbf{T}_2-\textbf{T}_1)}\end{array} & 0
\end{array}
\right).
\label{eq:HTB}
\end{equation}
The Hamiltonian is a $2 \times 2$ matrix because the honeycomb lattice is 
a Bravais lattice with a two-element base.
By diagonalizing this matrix we obtain the well-known energy bands of 
graphene
\begin{equation}
\label{eq:bandas}
\arraycolsep=3pt 
E_{\pm}(\textbf{k})=
\pm t\sqrt{2+2\cos\left(\sqrt{3}k_ya\right)+
4\cos \left(\frac{\sqrt{3}}{2}k_y a\right)\cos\left(\frac{3}{2}k_x a\right)}.
\end{equation}
There is a cone-like dispersion relation of Dirac particles around the 
Dirac points (DP), which are the two non-equivalent corners of the 
first Brillouin zone, \textbf{K} and \textbf{K'}.
This means that the low-energy excitations obey a conic dispersion 
relation and behave like Dirac fermions. 
Expressing $\textbf{k}=\textbf{K}+(q_x,q_y)$ with $\textbf{K}$ one of 
the DP, and in the long-wavelength approximation $q_x a, q_y a \ll 1$,
the Hamiltonian matrix takes the form
\begin{equation}
H^\textbf{K}_0(\textbf{q})=
  \frac{3ta}{2}
  \left(
    \begin{array}{c c}
      0          & q_x+iq_y \\
      q_x -i q_y & 0
    \end{array}
  \right) \,,
\end{equation}
which can be cast in the form of a 2D non-massive Dirac Hamiltonian,
\begin{equation}
  H^{\alpha}_0(\textbf{q}) = \hbar v_F (\alpha \sigma_x q_x - \sigma_y q_y).
\label{Hdirac}
\end{equation} 
Here the Fermi velocity  $v_F=3at/2\hbar$ is 300 times smaller than 
the speed of light, 
$\boldsymbol{\sigma}=(\sigma_x,\sigma_y)$ are the Pauli matrices, 
and we have associated the index $\alpha=\pm 1$ with each of the 
non-equivalent DP. 
The conic dispersion relation can be obtained by diagonalizing this matrix, 
being $E_\textbf{K}(\textbf{q})=\pm \hbar v_F q$ near each DP. 
The eigenstates of these Hamiltonians are spinors, whose two components are associated with the two elements of the lattice base. 
The low-energy states in cylindrical coordinates for a circular graphene
sheet of radius $r_0$ (we may assume that $r_0 \rightarrow \infty$ at the end
of the calculation) are \cite{rech-tra-ryc-bla-bee-mor,hen-gui,wun-sta-gui,
cse-pal-pet,per-rod-sta-lop}
\begin{equation}
\label{eq:estadosK}
\arraycolsep=3pt
\Psi_{m,\nu,K}^{v}=\frac{1}{\sqrt{2}}\left(\begin{array}{c}
\psi_{m+1,\nu} \\
i \psi_{m,\nu}
\end{array}\right),
\Psi_{m,\nu,K}^{c}=\frac{1}{\sqrt{2}}\left(\begin{array}{c}
\psi_{m+1,\nu} \\
-i \psi_{m,\nu}
\end{array}\right),
\end{equation}
near \textbf{K}, and
\begin{equation}
\label{eq:estadosK'}
\arraycolsep=3pt
\Psi_{m,\nu,K'}^{v}=\frac{1}{\sqrt{2}}\left(\begin{array}{c}
\psi_{m,\nu} \\
i \psi_{m+1,\nu}
\end{array}\right), 
\Psi_{m,\nu,K'}^{c}=\frac{1}{\sqrt{2}}\left(\begin{array}{c}
\psi_{m,\nu} \\
-i \psi_{m+1,\nu}
\end{array}\right),
\end{equation}
near \textbf{K'}, with 
\begin{equation}
  \psi_{m,\nu}(r,\theta) = \frac{N_{m,\nu}}{2\pi} J_m(q_{m,\nu}r) e^{im\theta}
\label{eq:psi},
\end{equation}
where $N_{m,\nu}=\sqrt{2}/r_0 J_{m+1}(x_{m,\nu})$ is a normalization constant, 
$\{m, \nu\}$ are integers, 
and $q_{m,\nu}=x_{m,\nu}/r_0$, 
where $x_{m,\nu}$ is the $\nu$th zero of $J_m(x)$.


\subsection{Pseudospin, angular momentum, and helicity}
\label{sec:pseudospin}

As is well known, the low-energy states of graphene are two-component 
spinors. 
These spinors should not be confused with the spin states of the electron
(spin is not included in our tight-binding model): 
rather, they are related to the physical lattice. 
More precisely, each component is associated with the relative amplitude of 
the Bloch function in each sublattice of the honeycomb lattice.
This degree of freedom is called pseudospin, for it plays a role in the 
Hamiltonian analogous to the one played by the regular spin in the Dirac Hamiltonian. 
It borrows the same SU(2) algebra but, unlike the isospin symmetry that 
connects protons and neutrons, pseudospin is actually an angular momentum, 
as has been shown by Mecklenburg and Regan.\cite{pseudospin1}
The common interpretation of this variable is that the pseudospin would be 
pointing up in $\hat{z}$ (outside the plane containing the graphene disk) 
in a state where all the electrons would be found in \textit{A} sites, 
while it would be pointing down in $\hat{z}$ if the electrons were 
located exclusively in the \textit{B} sublattice. 
Of course, in the eigenstates of the Hamiltonian, electrons are
distributed homogeneously between the two sublattices, so the pseudospin 
is contained in the $xy$-plane.
This fact can easily be corroborated for the states in  
Eqs.\ \eqref{eq:estadosK} and \eqref{eq:estadosK'}, since the mean value of 
the z-component of the operator associated with the pseudospin is 
identically zero for any of those states,
$\bra{m, \nu, \alpha} \sigma_z \ket{m, \nu, \alpha}= 0$.

Since our goal is to study the interaction of graphene with light 
carrying OAM, it is worth to examine the OAM of our electronic states. 
The $z$-component of the OAM operator is
$L_z=-i \hbar (\partial/\partial \theta) \mathbb{I}$,
where the identity operator acts on the pseudospin space.
Since the effective Hamiltonian near one of the DP has circular symmetry 
[it depends only on the absolute value of $(q_x,q_y)$], one could naively 
expect the Hamiltonian to commute with the OAM. 
However, this is not the case:
\begin{equation}
\left[H_0^{\alpha},L_z\right]
    = - i \hbar v_F \alpha \left(\sigma_x p_y + \sigma_y p_y\right).
\end{equation}
In order to construct a conserved angular momentum, we add the 
pseudospin to $L_z$, and define the total angular momentum as
\begin{equation}
\label{eq:Jz}
J_z^\alpha =L_z-\alpha \frac{\hbar}{2} \sigma_z.
\end{equation}
It can be easily shown that this operator does commute with the Hamiltonian.
In addition, our chosen basis states are eigenstates of this total angular 
momentum operator (near \textbf{K}),
\begin{eqnarray}
J_z \Psi_{m,\nu,K}&=&\left(\begin{array}{c}
\hbar (m+1) \psi_{m+1,\nu} \\
\hbar m (\pm i \psi_{m,\nu})
\end{array}\right)
- \frac{\hbar}{2} \left(\begin{array}{c}
\psi_{m+1,\nu} \\
-(\pm i) \psi_{m,\nu}
\end{array}\right)
=\left( m+ \frac{1}{2} \right) \hbar \Psi_{m,\nu,K}.
\end{eqnarray}
In a similar way we found for states near \textbf{K'} that their eigenvalue is 
also $j_z=(m+1/2)\hbar$. 
This means that for any state the total angular momentum is directly 
associated with the index $m$. 
In addition, since the mean value of the pseudospin is zero, the index $m$ 
can be associated with the mean value of the OAM too.

As an alternative interpretation, it is possible to see $L_z$ as an 
\textit{envelope} angular momentum, an operator that acts only on the 
envelope---macroscopic---part of the wave function, as can be seen from 
its diagonal form in  pseudospin space.
The total angular momentum we have defined contains thus information on 
both the macroscopic and the microscopic (pseudospin) parts of the 
wavefunction.

There is another operator that commutes with $H_0^{\alpha}$, as in the 
regular Dirac equation: the helicity operator, defined near \textbf{K} as
$\Sigma_K = (\hbar/2) \, \bm{\sigma}^* \cdot \hat{q}$, where 
$\hat{q}=(q_x,q_y)/(q_x^2+q_y^2)^{1/2}$.
The helicity is the component of the pseudospin in the direction of motion. 
The helicity is a constant of motion and is $+\hbar/2$ for conduction-band
states, and $-\hbar/2$ for valence-band states, and it does not depend on 
the DP. 
The main advantage of this quantum number is that, unlike the total 
angular momentum, it is a constant of motion that differentiates the states 
in the conduction band from those in the valence band.


\section{Interaction Hamiltonian}
\label{sec:Hint}

We first consider the vector potential of the TL beam in the Coulomb gauge, 
given by
\begin{eqnarray}
\mathbf{A}(\textbf{r},t)=
A_0 e^{i(q_z z-\omega t)}
\left[ \bm{\epsilon}_{\sigma} J_\ell(q r) e^{i \ell \theta} 
      - \sigma i \mathbf{\hat{z}} \frac{q}{q_z} J_{\ell+\sigma}(qr) 
        e^{i (\ell+ \sigma) \theta} \right] + c.c. ,
\label{eq:tl}
\end{eqnarray}
where $\bm{\epsilon}_{\sigma}= \mathbf{\hat{x}} + i \sigma \mathbf{\hat{y}}$ 
are the polarization vectors and $\sigma=\pm 1$. 
The radial profile of the beam is given by Bessel functions $J_\ell(qr)$ 
and $J_{\ell+\sigma}(qr)$.
Alternatively, one can work with Laguerre-Gaussian modes instead.\cite{LG}
(A comment on notation: note that $q_z$ and $q$ refer to the light beam, 
while $q_x$, $q_y$, and $q_{m,\nu}$, to the electrons.)
In order to describe the interaction between light and matter, we use 
the minimal-coupling Hamiltonian, obtained by the usual prescription 
$\textbf{p} \rightarrow \textbf{p}+e\textbf{A}$ in Eq.\ (\ref{Hdirac}), 
which guarantees local gauge invariance. 
By performing this substitution we get
\begin{eqnarray}
  H^{\alpha}&=& \hbar v_F (\alpha \sigma_x q_x - \sigma_y q_y) + 
  e v_F (\alpha \sigma_x A_x - \sigma_y A_y)\nonumber \\ 
  &\equiv & H^\alpha_0 + H^\alpha_{\text{int}},
\label{eq:Htot}
\end{eqnarray}
where we have already found the solutions for $H_0^{\alpha}$ in 
polar coordinates, given by Eqs.\ (\ref{eq:estadosK}) and 
(\ref{eq:estadosK'}).
Of course, the total Hamiltonian of the system is obtained by summing
over $\alpha$ the Hamiltonian of Eq.\ \eqref{eq:Htot}. 
Since graphene is a two-dimensional system, only components $x$ and $y$ 
of the electromagnetic field appear, and hence its $z$ component
is automatically eliminated from the interaction. 
Then, for a graphene disk placed at $z=0$, the vector potential \eqref{eq:tl}
effectively becomes
\begin{equation}
\label{tl2d}
  \textbf{A}(r,\theta,t)=A_0 (\mathbf{\hat{x}}+i \sigma \mathbf{\hat{y}}) 
  e^{-i\omega t} J_{\ell}(qr) e^{i\ell\theta} + c.c. .
\end{equation}
We define the following quantities
\begin{align}
  A^{(+)}=&A_0 e^{-i\omega t} J_\ell(qr) e^{i\ell\theta}, \\
  A^{(-)}=&A_0 e^{i\omega t} J_\ell(qr) e^{-i\ell\theta}, \nonumber
\label{eq:A+A-}
\end{align}
which can be associated with absorption and emission of one photon, 
respectively. 
Thus, we write the interaction Hamiltonian close to a Dirac point 
labelled by $\alpha$ and with a given polarization of the incoming 
light $\sigma$ as
\begin{equation}
  H^{\alpha,\sigma}_{\text{int}}= e v_F \left( 
  \begin{array}{c c}
    0 & 
    \begin{array}{c}
    (\sigma- \alpha)A^{(+)} + (\alpha+ \sigma)A^{(-)}
    \end{array} \\
    \begin{array}{c}
    (\alpha+ \sigma) A^{(+)} + (\alpha- \sigma) A^{(-)}
    \end{array}& 0
  \end{array}
  \right).
\label{eq:Hintalphasigma}
\end{equation}
As we described in previous sections, the spinor components are associated
with the two sublattices in the graphene structure.
Thus, each interaction Hamiltonian $H^{\alpha,\sigma}_{\text{int}}$,
having only off-diagonal elements, exchanges probability amplitude
between the two sublattices.
For example, for $\alpha=1$, i.e.\ near \textbf{K}, and $\sigma=+1$ one 
obtains
\begin{equation}
  H_{\text{int}}^{K,+}=2 e v_F
  \left(
  \begin{array}{c c}
      0 & A^{(-)} \\
      A^{(+)} & 0
  \end{array}
  \right) ,
\label{eq:HintK}
\end{equation}
and for $\alpha=-1$, i.e.\ near \textbf{K'}, and $\sigma=+1$ we have
\begin{equation}
  H_{\text{int}}^{K',+}=2 e v_F
  \left(
  \begin{array}{c c}
      0 & A^{(+)} \\
      A^{(-)} & 0
  \end{array}
  \right).
\label{eq:HintKprime}
\end{equation}
The role of the individual matrix elements in these Hamiltonians
becomes clearer when the rotating-wave approximation (RWA) applies. 
In the RWA, for a valence- to conduction-band transition, 
it is admisible to neglect $A^{(-)}$ in Hamiltonians 
\eqref{eq:HintK} and \eqref{eq:HintKprime}. 
Because of the resulting matrix form, it is clear that the action on 
an electron in a state close to $K$ is, in a sense, opossite to 
that on an electron in a state close to $K'$. 
While in each Dirac point $H_{\text{int}}^{\alpha,+}$ exchanges
pseudospin components, in K (K') the up(down)-component 
is eliminated in the final resulting state.
Then, from a microscopic point of view, an electron originally in 
a valence band state near a given Dirac point becomes "localized" 
in one sublattice.


\section{Interaction matrix elements}
\label{sec:em}

In order to calculate relevant observable quantities, we now obtain 
the light-matter interaction matrix elements for interband transitions.
In the long-wavelength approximation we neglect the intervalley 
transitions, i.e.\ terms that connect states near different Dirac points.
Hence, we need to calculate
\begin{equation}
\arraycolsep=3pt 
\label{eq:elmatriz}
\langle c,m',\nu',\alpha| H_{\text{int}}^{\alpha,\sigma} 
|v,m,\nu,\alpha \rangle = 
\int 
\Psi^{c\dagger}_{m',\nu',\alpha}(r,\theta) 
H^{\alpha,\sigma}_{\text{int}}(r,\theta)
\Psi^v_{m,\nu,\alpha}(r,\theta) d^2\textbf{r}.
\end{equation}
Using the definition of $\Psi$, Eqs.\ (\ref{eq:estadosK}) and 
(\ref{eq:estadosK'}), and the Hamiltonian, Eq.\ (\ref{eq:Hintalphasigma}), 
we can write the matrix elements in the following way
\begin{equation}
\label{eq:int+}
\langle c,m',\nu',\alpha| 
H_{\text{int}}^{\alpha,+} |
v m,\nu,\alpha,\rangle = 
2i e 
\alpha v_F \int \left( \psi^*_{m'+1,\nu'}A^{(-)}\Psi_{m,\nu}
                      + \psi^*_{m',\nu'}A^{(+)}\Psi_{m+1,\nu}
                \right) d\textbf{r},
\end{equation} 
\begin{equation}
\label{eq:int-}
\langle c,m',\nu',\alpha| 
H_{\text{int}}^{\alpha,-} 
|v,m,\nu,\alpha \rangle = 
-2i e \alpha v_F \int \left( \psi^*_{m',\nu'}A^{(-)}\Psi_{m+1,\nu}  
                            + \psi^*_{m'+1,\nu'}A^{(+)}\Psi_{m,\nu}
                      \right) d\textbf{r}.
\end{equation}
Each matrix element contains only two terms, one associated with the 
absorption and the other with the emission of a photon. 
In the RWA and defining
\begin{eqnarray}
I^{m',\nu'}_{m,\nu}&=& N_{m' \nu'} N_{m\nu} r_0^2 
                      \int_0^1 y J_{m'}(x_{m' \nu'}y) J_\ell(qr_0 y) 
                                 J_m(x_{m \nu}y) dy
\label{eq:defI}
\end{eqnarray}
we obtain
\begin{eqnarray}
\label{eq:RWA}
\arraycolsep=3pt 
  \bra{c,m',\nu',\alpha}
  H_{\text{int}}^{\alpha,+}
  \ket{v,m,\nu,\alpha} &\cong & 
    2\alpha i e v_F A_0 e^{-i\omega t} 
    I^{m+1+\ell,\nu'}_{m+1,\nu} 
    \delta_{m',m+(\ell+1)}, \nonumber \\
  \bra{c,m',\nu',\alpha}
  H_{\text{int}}^{\alpha,-}
  \ket{v,m,\nu,\alpha} &\cong & 
    -2i \alpha e v_F A_0 e^{-i\omega t} I^{m+\ell,\nu'}_{m,\nu} 
    \delta_{m',m+(\ell-1)}, \nonumber \\ 
  \bra{v,m',\nu',\alpha}
  H_{\text{int}}^{\alpha,+}
  \ket{c,m,\nu,\alpha} &\cong & 
    -2 i \alpha e v_F A_0 e^{i\omega t} I^{m-\ell,\nu'}_{m,\nu} 
    \delta_{m',m-(\ell+1)}, \nonumber \\ 
  \bra{v,m',\nu',\alpha}
  H_{\text{int}}^{\alpha,-}
  \ket{c,m,\nu,\alpha} &\cong & 
    2 i \alpha e v_F A_0 e^{i\omega t} 
    I^{m+1-\ell,\nu'}_{m+1,\nu} 
    \delta_{m',m-(\ell-1)}.
\end{eqnarray}
The transfer of angular momentum is signaled by the Kronecker's deltas, 
which relate the quantum numbers $m$ and $m'$ of the initial and final 
states.
This transferred angular momentum includes both OAM and spin.
This is to be expected given that both bands in graphene have 
the same microscopic angular momentum ($p$-type orbitals).
In this sense interband transitions in graphene are analogous to 
intraband transitions in regular semiconductors, as shown previously 
for intraband transitions in quantum rings.\cite{qui-ber,qui-tam-ber}


\section{Equations of motion for photo-excited electrons}
\label{sec:hem}

The graphene Hamiltonian in the absence of the electron-light 
interaction is
\begin{equation}
{\cal H}_0^\alpha = \hbar v_f 
             \sum_{m,\nu} q_{m,\nu} 
             \left( \hat{a}^{\alpha \dagger}_{c,m,\nu,\alpha} 
                    \hat{a}_{c,m,\nu}^\alpha - 
                    \hat{a}^{\alpha\dagger}_{v,m,\nu,\alpha} 
                    \hat{a}_{v,m,\nu}^\alpha\right) ,
\end{equation}
where $\hat{a}^{\alpha \dagger}_{\lambda,m,\nu}$ 
($\hat{a}_{\lambda,m,\nu}^\alpha$) 
are the creation (annihilation) operators for the electrons in the band 
tagged by the index $\lambda = c, v$ and in the valley $\alpha$. 
Considering only interband transitions, the electron-light interaction  
Hamiltonian is given by
\begin{eqnarray}
{\cal H}_{\text{int}}^{\alpha,\sigma} &=& 
    \sum_{m\nu,m'\nu'} 
      \left( 
        \langle c,m',\nu',\alpha| 
          H_{\text{int}}^{\alpha,\sigma} 
        |v,m,\nu,\alpha \rangle \, 
        \hat{a}^{\alpha \dagger}_{c,m',\nu'} 
        \hat{a}_{v,m,\nu}^\alpha \right. \nonumber \\ 
    &+& \left.  
      \langle v,m',\nu',\alpha| 
      H_{\text{int}}^{\alpha,\sigma} 
      |c,m,\nu,\alpha \rangle \,
      \hat{a}^{\alpha \dagger}_{v,m',\nu'} 
      \hat{a}_{c,m,\nu}^\alpha 
    \right).
\end{eqnarray}
In order to simplify the notation, hereafter we drop the $\alpha$ index 
whenever there is no ambiguity. 
The density matrix operator is
$\hat{\rho}_{\lambda',m',\nu'; \lambda,m,\nu}=
\hat{a}^{\dagger}_{\lambda',m',\nu'} 
\hat{a}_{\lambda,m,\nu}$. 
The equation of motion for this operator in the Heisenberg picture is:
\begin{eqnarray}
\label{eqn:heisen}
i\hbar \frac{d}{dt}\hat{\rho}_{\lambda', m', \nu';\lambda, m, \nu}
=\left[\hat{\rho}_{\lambda', m', \nu';\lambda, m, \nu},{\cal H}^{\sigma}\right].
\end{eqnarray}
This leads to the equations of motion for the density matrix operator,
in an analogous way as has been shown in Ref.\  [\onlinecite{qui2010}]. 
We consider the evolution of three types of operators: 
$\hat{\rho}_{c, m', \nu';c, m, \nu}$, 
$\hat{\rho}_{v, m', \nu'; v, m, \nu}$, and 
$\hat{\rho}_{v, m', \nu'; c, m, \nu}$
\begin{eqnarray}
\label{eq:eqmovcc}
i \hbar \frac{d}{dt} \hat{\rho}_{c, m', \nu';c, m, \nu} 
&=& 
\hbar v_F (q_{m, \nu} - q_{m', \nu'}) \hat{\rho}_{c, m', \nu';c, m, \nu} 
\nonumber \\  
&+& 
\sum_{n, \mu} \left(\langle c, m, \nu| 
H_{\text{int}}^{\sigma} 
|v, n, \mu \rangle \hat{\rho}_{c, m', \nu';v, n, \mu}  \right. \nonumber \\ 
&-& \left. \langle v, n, \mu| 
H_{\text{int}}^{\sigma} 
|c, m', \nu' \rangle \hat{\rho}_{v, n, \mu;c, m, \nu} \right) \,, \\
\label{eq:eqmovvv}
i \hbar \frac{d}{dt} \hat{\rho}_{v, m', \nu';v, m, \nu} 
&=& 
\hbar v_F (q_{m', \nu'} - q_{m, \nu}) \hat{\rho}_{v, m', \nu';v, m, \nu}
\nonumber\\ 
&+&
\sum_{n, \mu} \left(\langle v, m, \nu| 
H_{\text{int}}^{\sigma} 
|c, n, \mu \rangle \hat{\rho}_{v, m', \nu';c, n, \mu}  \right.\nonumber \\ 
&-& 
\left. \langle c, n, \mu| 
H_{\text{int}}^{\sigma} |v, m', 
\nu' \rangle \hat{\rho}_{c, n, \mu;v, m, \nu} \right) \,,\\
\label{eq:eqmovvc}
i \hbar \frac{d}{dt} \hat{\rho}_{v, m', \nu';c, m, \nu} &=& 
\hbar v_F (q_{m', \nu'} + q_{m, \nu}) 
\hat{\rho}_{v, m', \nu';c, m, \nu}\nonumber\\ 
&+&
\sum_{n, \mu} \left(\langle c, m, \nu| 
H_{\text{int}}^{\sigma} 
|v, n, \mu \rangle \hat{\rho}_{v, m', \nu';v, n, \mu}  \right.\nonumber\\  
&-& 
\left. \langle c, n, \mu| 
H_{\text{int}}^{\sigma} 
|v, m', \nu' \rangle \hat{\rho}_{c, n, \mu;c, m, \nu} \right).
\end{eqnarray}
These equations are, of course, also valid for the expectation values of 
the density matrix operators, taken over the initial state of the graphene. 
We denote these expectation values ${\rho}_{c, m', \nu';c, m, \nu}$, 
${\rho}_{v, m', \nu'; v, m, \nu}$ (called populations when $m=m'$ and 
$\nu=\nu'$, and intraband quantum coherences otherwise) and 
${\rho}_{v, m', \nu'; c, m, \nu}$ (interband coherences). 
Notice that in Eqs.\ \eqref{eq:eqmovcc}-\eqref{eq:eqmovvc} we keep the 
intraband coherences, which are essencial in the TL excitation process. 
These coherences are usually left out of the theory when the 
\textit{vertical-transition} approximation is made.


\subsection{Low-excitation regime}\label{sec:lowe}	

Due to the impossibility of solving analytically 
Eqs.\ \eqref{eq:eqmovcc}-\eqref{eq:eqmovvc} in all their generality, 
we will consider the case of low photoexcitation. 
In this case, an analytical perturbative approach is possible, which gives us 
the basic physical insight that we are looking for. 
We proceed in the following way, as in Ref.\ [\onlinecite{qui2010}]: we 
first solve Eq.\ \eqref{eq:eqmovvc} for ${\rho}_{v, m', \nu';c, m, \nu}$,
considering that in the quasi-equilibrium  and at temperatures much lower 
than the Fermi temperature for graphene, we have 
$\rho_{v, m',\nu';v, m,\nu}^{(0)}=\delta_{m,m'}\delta_{\nu,\nu'}$ and 
$\rho_{c, m',\nu';c, m,\nu}^{(0)}= 0$ (zeroth-order intraband elements). 
We then solve Eqs.\ \eqref{eq:eqmovcc} and \eqref{eq:eqmovvv} using the 
first-order solutions of Eq.\ \eqref{eq:eqmovvc}.
The equation of motion for the first-order interband polarization is
\begin{equation}
\label{eq:orden1}
\left[ i \hbar \frac{d}{dt} -  \hbar v_F (q_{m', \nu'} + q_{m, \nu}) \right] 
\rho_{v, m', \nu';c, m, \nu}^{(1)} =  
\langle c, m, \nu | H_{\text{int}}^{\sigma} |v, m', \nu' \rangle.
\end{equation}
and its solution in the RWA, for a monochromatic electromagnetic field 
turned on at $t=0$, is given by
\begin{eqnarray}
\label{eq:orden1cv}
\rho_{v, m', \nu'; c, m, \nu}^{(1)} &=& 
- \frac{1-e^{-i[v_F(q_{m, \nu}+q_{m',\nu'})- \omega]t}}
 {\hbar v_F (q_{m, \nu}+q_{m',\nu'}) - \hbar \omega} 
\langle c, m, \nu | H_{\text{int}}^{\sigma} 
|v, m', \nu' \rangle \nonumber \\ 
&\equiv & Y_{m, \nu; m', \nu'}(t) \langle c, m, \nu | 
H_{\text{int}}^{\sigma} |v, m', \nu' \rangle.
\end{eqnarray}
With this solution, we get the second order intraband coherence from 
Eqs.\ \eqref{eq:eqmovcc} and \eqref{eq:eqmovvv}
\begin{eqnarray}
\label{eq:orden2v}
    \rho_{v, m', \nu'; v, m, \nu}^{(2)}(t)
&=& \delta_{m', m} \delta_{\nu',\nu} -
    \frac{i}{\hbar} e^{-i v_F(q_{m',\nu'}-q_{m,\nu})t} \nonumber\\ 
     & \times &
     \sum_{n, \mu}
    \bra{v, m, \nu} H_{\text{int}}^{\sigma} \ket{c, n, \mu}
    \, \bra{c, n, \mu} H_{\text{int}}^{\sigma} \ket{v, m', \nu'}  \nonumber\\ 
  & \times &   \int_0^t dt' e^{i v_F (q_{m', \nu'}- q_{m,\nu})t}
    \left[
        Y_{n, \mu; m', \nu'}(t) -
        Y_{m, \nu; n, \mu}^*(t)
    \right]\,,
\end{eqnarray}
for the valence band, and
\begin{eqnarray}
\label{eq:orden2c}
    \rho_{c, m',\nu'; c, m, \nu}^{(2)}(t)
&=&-
    \frac{i}{\hbar} e^{-i v_F(q_{m,\nu}-q_{m',\nu'})t}\nonumber \\
& \times &     \sum_{n, \mu}
    \bra{c, m, \nu} H_{\text{int}}^{\sigma} \ket{v, n, \mu}
    \, \bra{v, n, \mu} H_{\text{int}}^{\sigma} \ket{c, m', \nu'} \nonumber\\ 
  & \times &   \int_0^t dt' e^{i v_F (q_{m, \nu}- q_{m',\nu'})t}
    \left[
        Y_{n, \mu; m', \nu'}^*(t) -
        Y_{m, \nu; n, \mu}(t)
    \right]\,.
\end{eqnarray}
for the conduction band. 
As an example, we show the populations for a TL beam with right-handed 
circular polarization:
\begin{align}
\label{eq:popv}
n^{(2)}_{v, m, \nu}(t)=&1-\frac{8e^2 v_F^2 A_0^2}{\hbar^2}
 \sum_{\mu} 
\frac{\left(I^{m+1+\ell,\mu}_{m+1,\nu}\right)^2
 \left\lbrace 1- \cos [(v_F(q_{m,\nu}+q_{m+(\ell+1),\mu})-\omega)t] 
 \right\rbrace}{\left[v_F(q_{m,\nu}+q_{m+(\ell+1),\mu})-
\omega \right]^2} \,, \\
\label{eq:popc}
n^{(2)}_{c, m, \nu}(t)=&\frac{8e^2 v_F^2 A_0^2}{\hbar^2}  
\sum_{\mu} \frac{\left(I^{m-\ell,\mu}_{m,\nu}\right)^2
\left\lbrace 1- \cos [(v_F(q_{m,\nu}+q_{m-(\ell+1),\mu})-\omega)t] 
\right\rbrace}
{\left[v_F(q_{m,\nu}+q_{m-(\ell+1),\mu})-\omega \right]^2} \,.
\end{align}
Near resonance, the populations evolve slowly, as expected. 
Also, the angular momentum of the intermediate state differs from that of 
the initial state by $\ell \pm 1$, the total angular momentum of the photon.

\begin{figure*}
\centerline{\includegraphics[width=\linewidth]{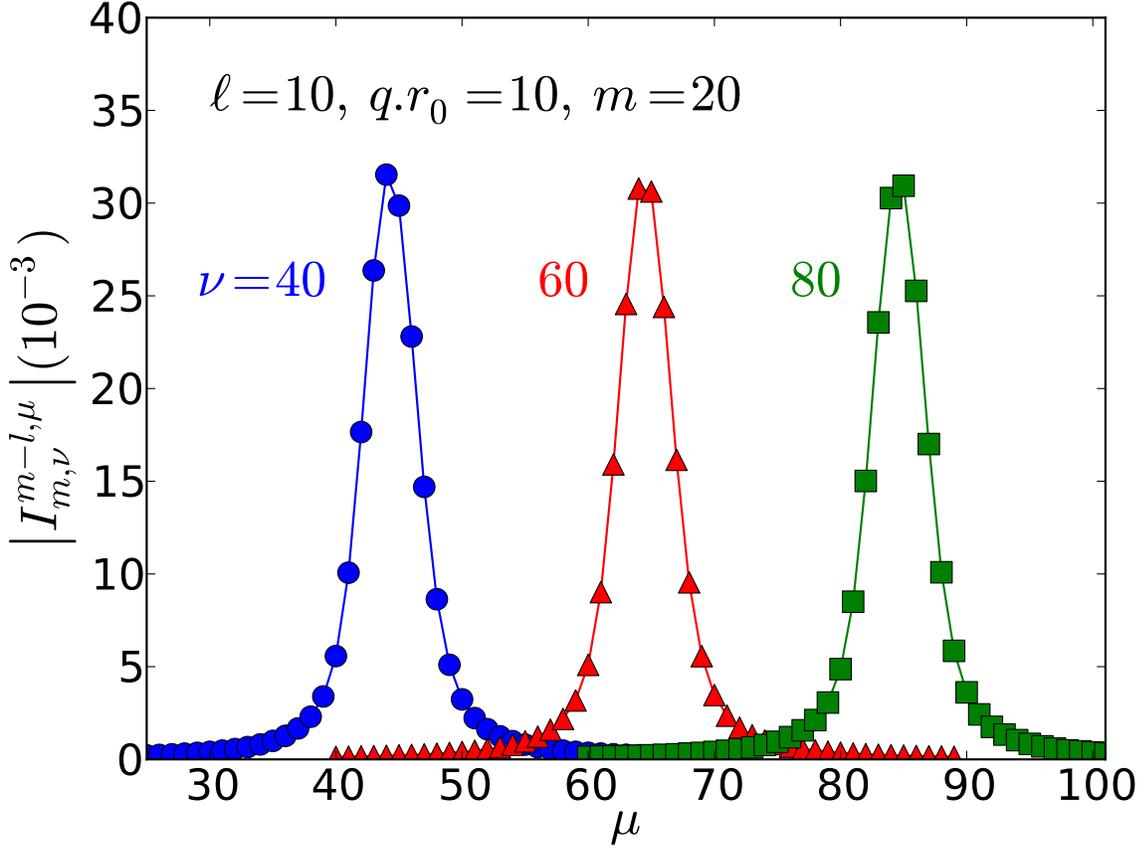}}
\caption{Absolute value of the integral proportional to the matrix elements 
of the interaction of twisted light with graphene, which determines the 
population of the conduction-band excited states [see Eqs.\ \eqref{eq:defI} 
and  \eqref{eq:popc}], for three different values of the quantum number
$\nu$ of the final state.}
\label{fig:various_nu}
\end{figure*}

The populations' functional forms reflect, in part, what is expected based 
on general ideas of optical excitation. 
However, they are affected by the parameter dependence of the integral 
$I^{m',\nu'}_{m,\nu}$, which is proportional to the light-matter matrix 
elements and carries information about the twisted character of the 
electromagnetic field. 
Therefore, to gain more insight into the population kinetics, we calculate 
[using Eq.\ \eqref{eq:defI}] and discuss some examples of the integrals 
$I^{m-\ell,\mu}_{m,\nu}$, which contribute to the population of 
the conduction-band states according to Eq.\ \eqref{eq:popc}. 
We start by fixing the system (TL beam and graphene disk) parameters $\ell$ 
and $q.r_0$ and inspect which matrix elements are relevant for the population 
of a state $(c,m,\nu)$ of the conduction band.
Since $I^{m-\ell,\mu}_{m,\nu}$ enters the population through a sum over its 
$\mu$ parameter, in Fig.\ \ref{fig:various_nu} we plot three different 
curves of $|I^{m-\ell,\mu}_{m,\nu}|$, for $\nu=40, 60, 80$, as a funcion 
of $\mu$ (with $\ell=10$, $q.r_0=10$, and $m=20$). 
For this choice of parameters, we notice that the curves have a fairly
well-behaved bell shape, and that the peak value occurs close to $\mu=\nu$, 
slightly shifted to the right.
By plotting similar curves for various values of $m$, one can easily
conclude that they are mainly insensitive to the value of $m$.
 
\begin{figure*}
\centerline{\includegraphics[width=\linewidth]{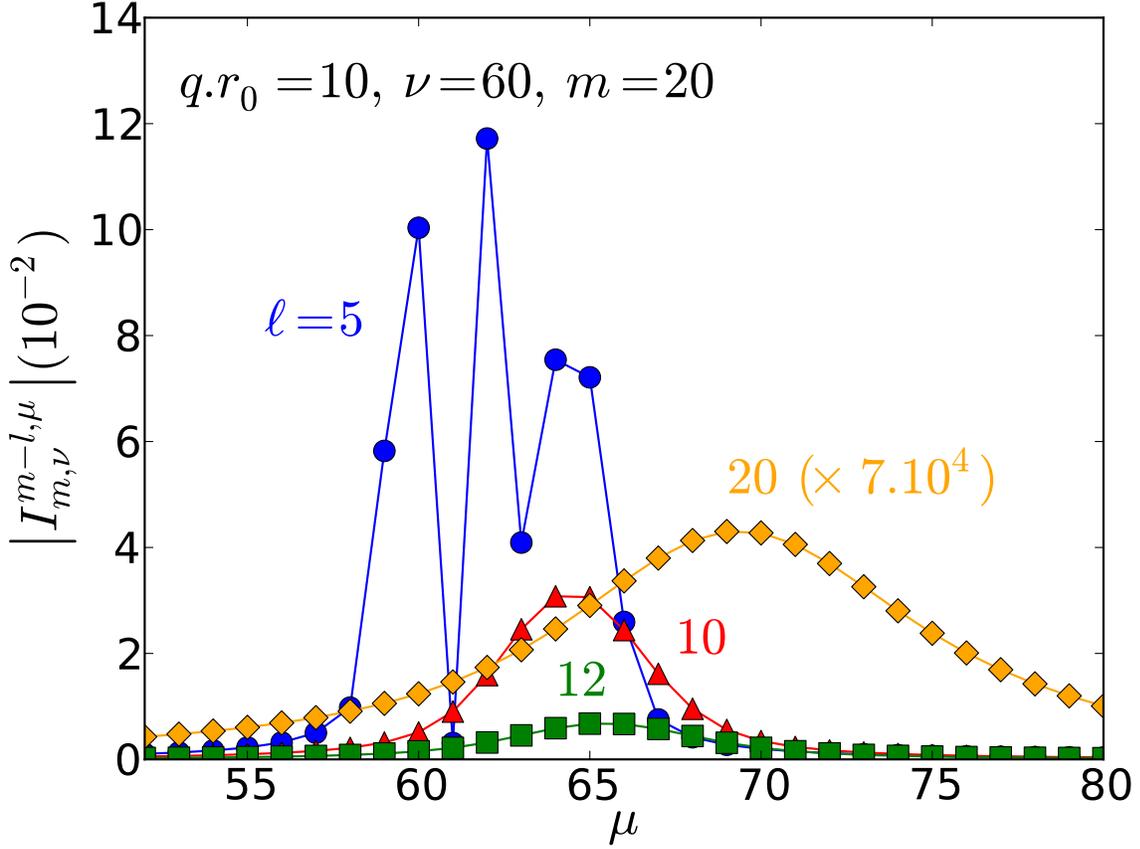}}
\caption{Same as Fig.\ \ref{fig:various_nu}, for four different values of 
the twisted-light beam's orbital angular momentum, $\ell$.}
\label{fig:various_ell}
\end{figure*}

\begin{figure*}
\centerline{\includegraphics[width=\linewidth]{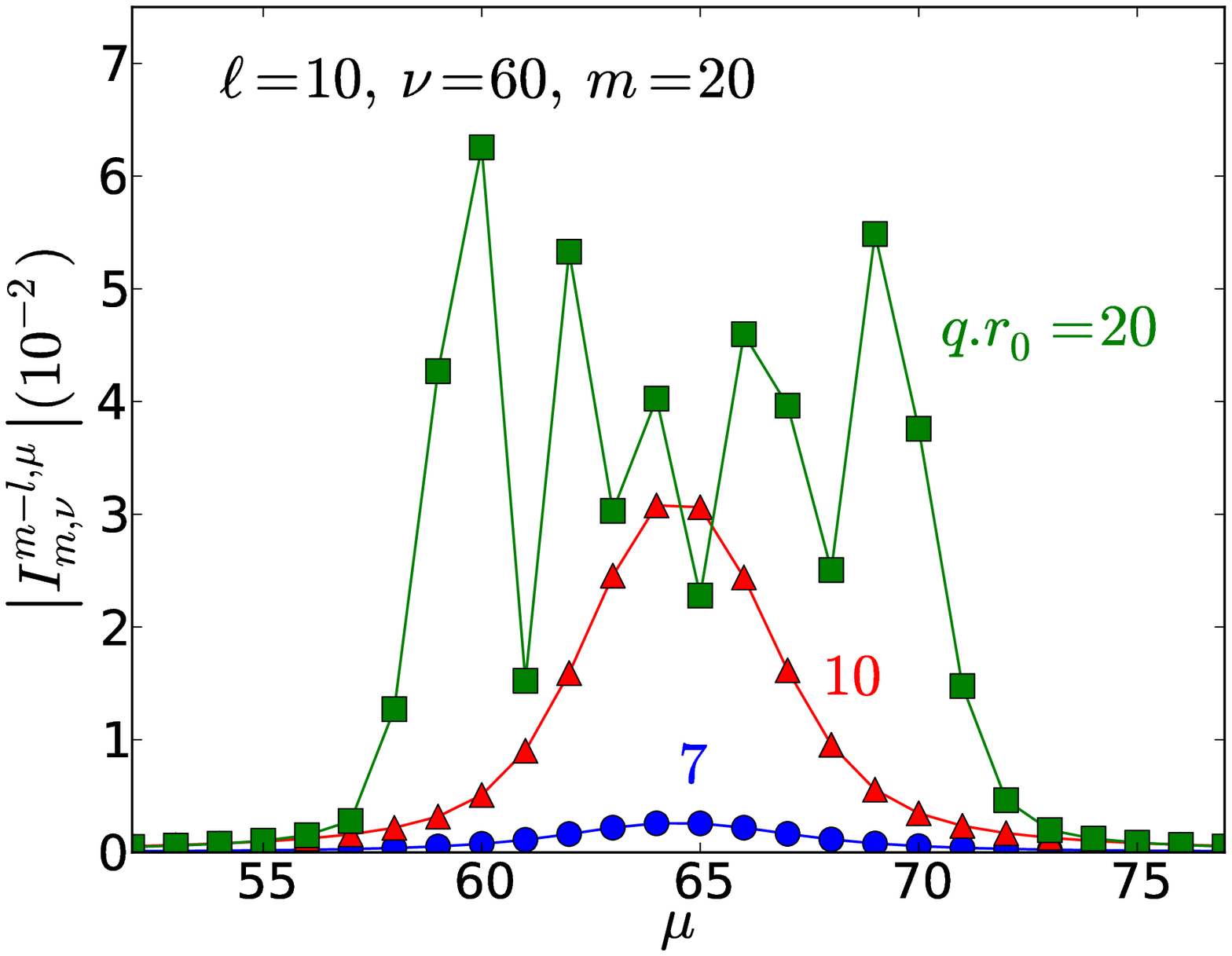}}
\caption{Same as Fig.\ \ref{fig:various_nu}, for three different values of 
the system (twisted-light beam and graphene disk) parameter $q.r_0$.}
\label{fig:various_qr0}
\end{figure*}

Now we can vary the light-beam and graphene disk parameters.
In Fig.\ \ref{fig:various_ell}, we plot $|I^{m-\ell,\mu}_{m,\nu}|$ 
for $\ell=5, 10, 12, 20$ as a funcion of $\mu$ (with $q.r_0=10$, 
$\mu=60$, and $m=20$).
We notice that the curves have a well-defined bell shape for
the larger values of $\ell$, but not for $\ell=5$.
Furthermore, for the former cases, we see that the bell widens 
as $\ell$ increases, and at the same time, it rapidly decreases 
in amplitude.
For (small) $\ell=5$, the nice bell shape is lost and many values
of $\mu$ contribute to the population with similar weight.
All the cases without a sharp bell-like curve share an important
characteristic of the population kinetics: many valence-band
states contribute to the population of a given conduction-band state, 
and therefore a clear picture of Bloch oscillations is lost. 
Finally, we vary the parameter $q.r_0$, which combines the radial
dependence of the beam with the radius of the graphene disk.
A large value of $q.r_0$ means that the light field makes many
oscillations in the radial direction inside the material disk.
In Fig.\ (\ref{fig:various_qr0}), we plot $|I^{m-\ell,\mu}_{m,\nu}|$ 
for $q.r_0=7, 10, 20$ as a funcion of $\mu$ (with $\ell=10$, 
$\mu=60$, and $m=20$).
It is seen that for $q.r_0=7, 10$ one obtains a clean bell shape,
while for $q.r_0=20$ the picture is similar to that of the $\ell=5$ 
(and $q.r_0=10$) case in Fig.\ \ref{fig:various_ell}.
We interpret that this kind of behavior appears whenever $q.r_0 > \ell$
and therefore the light field "oscillates more'' in the radial coordinate 
than in the angular one, in which case even an approximate selection rule
for the quantum number $\nu$ is lost.

\section{Evolution of physical magnitudes}
\label{sec:epm}

The modification of the electronic state of graphene by twisted light 
can be described in terms of the evolution of key physical quantities.
In this Section we find the expressions for the photocurrents and the orbital 
angular momentum, using the lowest-order contributions to the density
matrix obtained above.


\subsection{Transfer of angular momentum}
\label{Sec:transfOM}

In Sec.\ \ref{sec:pseudospin} we defined operators for the orbital and 
total angular momentum. 
In second quantization formalism and taking the expectation value 
over the ground state of graphene, the total angular momentum, which is 
diagonal in our basis, is given by the simple expression
\begin{equation}
\label{eq:Jz2}
J_z(t)=2 \sum_{\lambda, m, \nu}  
         \left( m + \frac{1}{2} \right) 
         n_{\lambda, m, \nu}^{(2)}(t),
\end{equation}
where the factor 2 comes from the valley degeneracy. 
$J_z(t)$ depends only on the populations, and thus
evolves slowly with a magnitude proportional to amplitude of the 
electromagnetic field squared.

On the other hand, we get for the OAM in operator form a slightly more complicated expression
\begin{eqnarray}
L_z(t)&=& \sum_{\alpha, m, \nu} 
          \left(m+\frac{1}{2}\right) \hbar 
          \left[ n^{\alpha (2)}_{c, m,\nu}(t)+n^{\alpha(2)}_{v,m,\nu}(t)\right] 
          \\ 
&+& \frac{\hbar}{2} \sum_{m \nu} \sum_{\lambda \neq \lambda'} 
\left[\rho^{K(1)}_{\lambda', m, \nu; \lambda, m, \nu}(t) - 
      \rho^{K'(1)}_{\lambda', m, \nu; \lambda, m, \nu}(t) \right].
\end{eqnarray}
However, this expression greatly simplifies when we take the expectation 
value over an initial state, for 
$\rho^{K(1)}_{\lambda' m \nu, \lambda m \nu} \propto 
\delta_{m,m+(\ell\pm 1)} = 0$. 
Then, the expectation value of the OAM coincides with that of the 
total angular momentum.


\subsection{Diamagnetic contribution to the angular momentum}

In Sec.\ \ref{sec:Hint}, we included the light field by the substitution 
$\textbf{p} \rightarrow \textbf{p} + e \textbf{A}$.
In the angular momentum operator, this substitution introduces the 
so-called diamagnetic term  
$L_z^{(dia)}=e\left(\textbf{r} \times \textbf{A} \right)_z$,
which we now consider. 
Its single-particle form is
\begin{equation}
\label{eq:Lzdia2}
L_z^{(dia)}=e i \sigma A_0 r J_\ell(qr) 
            \left[ e^{-i \omega t} e^{i (\ell+\sigma)\theta}-
                   e^{ i \omega t} e^{-i(\ell+\sigma)\theta}
            \right].
\end{equation}
This operator is of first-order in $A_0$.
The expectation value of any operator can be expressed, in our scheme, 
as a sum of terms proportional to the populations and coherences, each 
multiplied by the appropriate matrix element of the single-particle operator. 
Since populations are of second-order in $A_0$, the corresponding term in 
$\langle L_z^{(dia)}\rangle$ is proportional to $A_0^3$ and is thus 
neglected in our second-order, low-excitation treatment.
Instead, the coherence terms in $\langle L_z^{(dia)} \rangle$ are of 
order $A_0^2$, and should be kept.
The latter terms are especially relevant since the coherence terms
of the paramagnetic angular momentum, which would be of first order,
actually vanish here (see Sect.\ \ref{Sec:transfOM}).
The matrix elements of $L_z^{(dia)}$ are given by
\begin{align}
\bra{\lambda', m', \nu', K} L_z^{(dia)} \ket{\lambda, m, \nu, K} 
&= e i \sigma A_0 \left[ \delta_{m',m+(\ell+\sigma)} e^{-i \omega t} - 
\delta_{m',m-(\ell+\sigma)} e^{i \omega t} \right] \times \nonumber\\
&\times \int  \left[ J_{m'+1,\nu' } (q_{m'+1,\nu' }r) 
                J_{m+1,\nu}(q_{m+1,\nu }r)J_\ell(qr) \right. \nonumber \\ 
&\left. - J_{m',\nu' } (q_{m',\nu' }r) J_{m,\nu}(q_{m,\nu}r)
          J_\ell(qr)\right] r^2 dr .
\end{align}
The second-quantized form of $L_z^{(dia)}$ requires a summation over all 
possible values of $m,m',\nu$ and $\nu'$. 
The summation over $m'$ results in only one term, and the summation over 
$m'$ can be thought in the following way:
\begin{align}
\label{eq:forallm}
&\sum_{\forall m} \int dr r^2 [ J_{m\pm(\ell+\sigma)+1,\nu' } (q_{m\pm(\ell+\sigma)+1,\nu' }r) J_{m+1,\nu}(q_{m+1,\nu }r)J_\ell(qr)] \nonumber \\
&\approx \sum_{\forall m} \int dr r^2 [ J_{m\pm(\ell+\sigma),\nu' } 
(q_{m\pm(\ell+\sigma)+1,\nu' }r) J_{m,\nu}(q_{m,\nu }r)J_\ell(qr)].
\end{align}
This expression is only approximately accurate, since we are not actually 
summing over all $m$, but the two members of Eq. \eqref{eq:forallm} differ 
only in two terms, and we will neglect them in this work.


\subsection{Induced photocurrents}

In this Section we obtain the photoinduced currents produced by the 
irradiation with TL. 
Their first-quantized form is obtained by writing the interaction 
Hamiltonian as 
$H^\alpha_{\text{int}}=-\bm{\jmath}^{\alpha} \cdot \textbf{A}$, thus:
\begin{equation}
  \bm{\jmath}^\alpha = - e v_F (\alpha \sigma_x, - \sigma_y).
\label{eq:j1c}
\end{equation}
In second quantization notation, the density current will be given by:
\begin{equation}
  \bm{\jmath}^\alpha =e v_F \hat{\Psi}^{\alpha\dagger}(\textbf{r},t) 
  (-\alpha \sigma_x,\sigma_y) \hat{\Psi}^{\alpha}(\textbf{r},t),
\label{j2c}
\end{equation}
where $\hat{\Psi}^{\alpha}$ are the field operators of the system. 
This result is consistent with the current obtained from the Dirac 
Hamiltonian. 
Given the symmetries of this problem, and in order to find a relation 
between the induced currents and the transfer of angular momentum, 
we write the currents in cylindrical coordinates, and separate its 
$\hat{r}$ and $\hat{\theta}$ components. 
By writing the field operators in terms of creation and annihilation
operators, we get
\begin{eqnarray}
 \jmath^{\alpha}_{r}(\textbf{r},t) = - 
 e v_F \sum_{\lambda, \lambda'} \sum_{m, m'} \sum_{\nu, \nu'}
 \Psi^{\alpha \dagger}_{\lambda, m, \nu}(\textbf{r})
 \left( \alpha \sigma_x \cos \theta - \sigma_y \sin \theta \right) 
 \Psi^{ \alpha}_{\lambda', m', \nu'}(\textbf{r}) 
 \rho^{\alpha}_{\lambda, m, \nu;\lambda', m', \nu'}(t) , \\
 \jmath^{\alpha}_{\theta}(\textbf{r},t) =
 - e v_F \sum_{\lambda, \lambda'} \sum_{m, m'} \sum_{\nu, \nu'}
 \Psi^{\alpha \dagger}_{\lambda, m, \nu}(\textbf{r}) 
 \left( -\alpha \sigma_x \sin \theta - \sigma_y \cos \theta \right) 
 \Psi^{ \alpha}_{\lambda', m', \nu'}(\textbf{r}) 
 \rho^{\alpha}_{\lambda, m, \nu; \lambda', m', \nu'}(t).
\end{eqnarray}
We next write separately the interband coherence contribution from 
the intraband coherence and population contributions.

An expression for the interband coherence contribution can be obtained 
using the functional form of  
$\rho^{\alpha (1)}_{v,m,\nu; c, m', \nu'}$ 
[Eq.\ (\ref{eq:orden1cv})] 
and the matrix elements in the RWA [Eq.\ \eqref{eq:RWA}]. 
We show the results for left-handed polarized light ($\sigma=1$). 
The $\hat{r}$ component reads
\begin{align}
&\jmath_{r}^{\alpha(coh)}(\textbf{r},t) = 
-4 A_0e^2v_F^2 \cos[(\ell+1)\theta]  \nonumber \\
& \times \sum_{m,\nu,\nu'} \frac{N_{m+1,\nu}N_{m+\ell+1,\nu'}}{2\pi}
                           J_{m+1}(q_{m+1,\nu}r)
                           J_{m+\ell+1}(q_{m+\ell+1,\nu'}r) \nonumber  \\
& \times  I^{m+1+\ell,\nu'}_{m+1,\nu}(q,\ell) 
          \left[e^{-i \omega t} Y_{(m+\ell+1),\nu';m,\nu}(t) + 
                e^{i \omega t} Y^*_{(m+\ell+1),\nu';m,\nu}(t)  \right]. 
\end{align}
For the $\hat{\theta}$ component we get
\begin{align}
&\jmath_{\theta}^{\alpha (coh)}(\textbf{r},t)=
4 A_0e^2v_F^2 \sin[(\ell+1)\theta] \nonumber \\ 
& \times\sum_{m,\nu,\nu'} \frac{N_{m+1,\nu}N_{m+\ell+1,\nu'}}{2\pi}
                          J_{m+1}(q_{m+1,\nu}r)
                          J_{m+\ell+1}(q_{m+\ell+1,\nu'}r) \nonumber \\
&\times  I^{m+1+\ell,\nu'}_{m+1,\nu}(q,\ell) 
         \left[e^{-i \omega t} Y_{(m+\ell+1),\nu';m,\nu}(t) + 
               e^{i \omega t} Y^*_{(m+\ell+1),\nu';m,\nu}(t)  \right].
\end{align}
We note that the $\hat{\theta}$ component is identically zero when 
$\ell = -\sigma$, which shows that if the photon carries no total angular 
momentum the induced currents do not rotate around the beam's axis.
On the other hand, the $\hat{r}$ component remains finite for all values
of $\ell$.

For the population and intraband coherences, the lowest non-trivial order 
does not depend on the valley. 
Noticing that 
$\rho^{(2)}_{\lambda, m, \nu;\lambda, m', \nu'} \propto \delta_{m,m'}$ 
for all $\lambda, \nu, \nu'$, 
the $\hat{r}$ component will be identically zero at second order 
in $A_0$, since
\begin{equation}
\jmath_r^{(pop)}(\textbf{r},t) \propto \sin[(m'-m)\theta] 
                                               \delta_{m',m}=0,
\end{equation}
and the $\hat{\theta}$ component does not depend on the azimuthal 
coordinate, and is given by
\begin{equation}
\jmath_{\theta}^{(pop)}(\textbf{r},t)=
2ev_F \sum_{m,\nu,\nu'} \frac{N_{m+1,\nu}N_{m,\nu'}}{2\pi}
J_{m+1}(q_{m+1,\nu}r) J_{m}(q_{m,\nu'}r) 
\left( \rho^{(2)}_{v, m, \nu; v, m, \nu'}(t) - 
       \rho^{(2)}_{c, m, \nu; c, m, \nu'}(t) \right).
\end{equation}
Here, the OAM of the light, $\ell$, appears implicitely through the 
population and intraband coherences [see Eqs.\ \eqref{eq:popv}
and \eqref{eq:popc}].
Finally, its worth mentioning that even in the case with $\ell=0$ 
(i.e.\ when the light does not carry OAM and it is just a regular beam 
with a radial Bessel dependence instead of Gaussian), there still is a 
transfer of angular momentum to the electrons, evidenced by a photocurrent 
that rotates around the beam's axis. 
This can be seen as a transfer of spin angular momentum of the photons 
to the orbital angular momentum of the electrons.
This does not happen in the case of interband transitions in 
semiconductors,\cite{qui2010} where the matrix elements are proportional 
to $\delta_{m',m\pm \ell}$. 
Our result, however, agrees with the findings of the study of intraband transitions in semiconductors, for the reasons explained in a previous 
section. 


\section{Conclusions}

We studied theoretically the interaction of graphene with twisted light.
As a starting point, using the tight-binding states of electrons in
graphene near the Dirac points, we determined the light-matter interaction
matrix elements.

In the RWA, for a valence- to conduction-band transition, we found that
the action of the light-matter Hamiltonian on an electron in a state close 
to $K$ is, in a sense, opposite to that on an electron in a state close 
to $K'$.
The pseudospin components are exchanged in each Dirac point, 
and in addition, in $K$ ($K'$) the up(down)-component is eliminated 
in the final resulting state. 

We examined the dynamics of the photoexcitation process by posing the
equations of motion of the density matrix, and obtaining the time
evolution of the angular momentum of the photoexcited electrons
and their associated photocurrents.
The transfer of spin and orbital angular momentum from light to the 
electrons, is more akin here to the case of intraband than of interband
transitions in semiconductors.
This is due to the fact that the two relevant energy bands of graphene 
originate from the same atomic orbitals. 

\acknowledgments
We acknowledge financial support from the University of Buenos Aires
through the program UBACYT 2011-2014 and Beca Est\'imulo for M.B.F.
Figures \ref{fig:various_nu}, \ref{fig:various_ell}, and \ref{fig:various_qr0} 
were made with the graphics package matplotlib.\cite{hun}



\end{document}